\documentclass[12pt]{article}

\usepackage{amssymb,graphicx}

\usepackage{epsf}
\usepackage{graphicx,epsfig}
\usepackage{amsfonts}
\usepackage{amssymb}
\usepackage{cite}

\def\bx{{\bf x}}

\def\CO{{\cal O}}

\def\mpl{M_{\rm Pl}}
\def\half{\frac{1}{2}}

\makeatletter
\renewcommand\section{\@startsection {section}{1}{\z@}%
                                 {-3.5ex \@plus -1ex \@minus -.2ex}
                                   {2.3ex \@plus.2ex}%
                                   {\normalfont\large\bfseries}}
\renewcommand\subsection{\@startsection{subsection}{2}{\z@}%
                                   {-3.25ex\@plus -1ex \@minus -.2ex}%
                                     {1.5ex \@plus .2ex}%
                                     {\normalfont\bfseries}}
\renewcommand\subsubsection{\@startsection{subsubsection}{3}{\z@}%
                                   {-3.25ex\@plus -1ex \@minus -.2ex}%
                                     {1.5ex \@plus .2ex}%
                                     {\normalfont\itshape}}
\makeatother

\newcommand{\Letter}{
\setlength{\textwidth}{16.5cm}
   \setlength{\textheight}{22.6cm}
    \hoffset=-0.5in
\voffset=-2.1cm }

\Letter

\begin{document}
\newcommand{\be}{\begin{equation}}
\newcommand{\ee}{\end{equation}}
\newcommand{\bea}{\begin{eqnarray}}
\newcommand{\eea}{\end{eqnarray}}
\newcommand{\barr}{\begin{array}}
\newcommand{\earr}{\end{array}}

\thispagestyle{empty}
\begin{flushright}
\parbox[t]{1.5in}{MIT-CTP-3961}
\end{flushright}

\vspace*{0.3in}

\begin{center}
{\large \bf Fine-Tuning in DBI Inflationary Mechanism}

\vspace*{0.5in} {Xingang Chen}
\\[.3in]
{\em Center for Theoretical Physics 
\\ Massachusetts Institute of Technology, Cambridge, MA 02139 }
\\[0.3in]
\end{center}

\begin{center}
{\bf
Abstract}
\end{center}
\noindent
We show a model-independent 
fine-tuning issue in the DBI inflationary mechanism. DBI
inflation requires a warp factor $h$ small enough to sufficiently slow
down the inflaton. On the other hand, 
the Einstein equation in extra dimensions under the inflationary
background deforms the warp space in the IR
side. Generically these two locations coincide with each other,
spoiling the DBI inflation. The origin and tuning of this
``$h$-problem'' is closely related,  through the
AdS/CFT duality, to those of the well-known 
``$\eta$-problem'' in the slow-roll inflationary mechanism.

\vfill

\newpage
\setcounter{page}{1}

\section{Introduction}
One important motivation for the DBI (Dirac-Born-Infeld) inflation
\cite{Silverstein:2003hf,Chen:2004gc} is
to circumvent the $\eta$-problem in the slow-roll inflation
\cite{Copeland:1994vg}. Namely,
the required inflaton mass-squared for the slow-roll inflation, 
$\lesssim$ $\CO(0.01 H^2)$, is generically 
corrected to $\CO(H^2)$ due to the backreaction of the inflationary
background.
This leads to a potential too steep to support the slow-roll
inflation. It is proposed that warped space can provide a
speed-limit which holds the inflaton near the top of a potential
even if the potential is steep. Indeed, given such a warped space, it
is shown that the
inflation can happen, in different situations, with a very steep
potential (typically
$m^2 \gg H^2$) \cite{Silverstein:2003hf} or a potential of
generic shapes (typically $m^2\sim\CO(H^2)$) \cite{Chen:2004gc}. See
Ref.~\cite{Bean:2007eh} for a summary.
The importance of the warped space for this mechanism is
obvious. 

In this paper we discuss a subtlety involved in the construction of
such a warped space. At the level of the 4d effective field theory, 
the warping effect shows up in the kinetic term of the inflaton $r({\bf
  x}, t)$ through the warp factor $h(r)$,
\bea
h^4 \sqrt{1+h^{-4} g^{\mu\nu} \partial_\mu r \partial_\nu r} ~,
\label{DBIkinetic}
\eea
and it seems that we can take whatever $h(r)$ we like.
A typical example is the AdS$_5$ space with a
scale $R$,
\bea
h(r) = \frac{r}{R} ~.
\label{hAdS}
\eea
In oder to provide a speed-limit small enough for the inflation to
happen, $h$ is required to be smaller than a critical
value $h_0$.

On the other hand, the warped space lies in the internal field space,
or the extra dimensions. In order to have a consistent UV completion
of the DBI inflation, such as in terms of brane inflation
\cite{HenryTye:2006uv}, the warp
factor is supplied by a metric
\bea
ds^2= h(r)^2 (-dt^2+ a(t)^2 d {\bf x}^2) + h(r)^{-2} dr^2 ~,
\label{5dmetric}
\eea
where $a(t)$ is the scale factor of the 4d inflationary background.
This metric has to satisfy
Einstein's equation in the extra dimensions. As we will see, due to the
backreaction of the 4d inflationary background, 
the warp factor $h$ has to be
deformed away from (\ref{hAdS}) near a critical value. 
Generically, this value coincides with what is required for the
inflation, $h_0$, and the resulting metric leads to a huge inflaton
probe backreaction. So tuning is necessary.

We comment that this tuning issue refers to the {\em mechanism}
of the DBI inflation \cite{Silverstein:2003hf,Chen:2004gc}, 
which provides the bases for various
observational consequences
\cite{Alishahiha:2004eh,Chen:2005ad}.
With this mechanism achieved, there may be other
tuning issues involving model parameters when data comparison is
made \cite{Alishahiha:2004eh,Chen:2005ad}.
To distinguish from these secondary issues, we
call the problem discussed here the ``$h$-problem''.

Maldacena first emphasized, a few years ago, the importance of the
backreaction from the Hubble expansion 
to the DBI inflationary mechanism \cite{Malda}.
Generally we expect a warped space with a very small minimum warp
factor to be deformed in the IR side by the Hubble expansion, so the
location where the deformation starts to be significant is crucial.
So far there have been several
kinds of such backreactions discussed in the literature
\cite{Chen:2005ad,Frey:2005jk,Chen:2006ni,McAllister:2007bg,Becker:2007ui}.
Particles such as strings can be created in the IR side 
by the spatial expansion because their mass scales are
redshifted. Only when
the density of these particles
exceeds certain critical value, its backreaction shortens the
warped throat; but this still leaves
a considerable portion of warped space for DBI inflation
\cite{Chen:2005ad}. Various moduli that stabilize the IR warp factor
in the flux compactification will fluctuate in the inflationary
background and this also shortens the throat
\cite{Frey:2005jk,Chen:2006ni}. The analyses along this line so far
has not
been explicit enough for our purpose due to technical difficulties
in solving equations of motion with large warping, and also it is likely to
be model-dependent. The throat can also be deformed due to the
mediation of the supersymmetry
breaking \cite{McAllister:2007bg}, such effects are model-dependent of
the details of the supersymmetry breaking. 
As mentioned, in this paper we explain
the most general and significant source of the
backreaction to the DBI inflationary mechanism,
based on the Einstein equations for the metric.
The importance of identifying such a source has an analogue in the
slow-roll inflation, where the most general source of the
$\eta$-problem is identified to be the canonical inflaton dependence in the
Kahler potential \cite{Copeland:1994vg}, and we can therefore look for
other model-dependent backreactions to tune away this
problem. Similarly for DBI inflation, other model-dependent
backreaction aspects mentioned above,
such as the particle
creation, moduli shifting and the supersymmetry breaking, may provide
the counter-sources to tune
away the $h$-problem, as we will discuss.

We will also see that the origin and tuning of the $h$-problem in DBI
inflation is closely
related to those of the $\eta$-problem in slow-roll inflation 
through the AdS/CFT duality.
Hopefully both the explicit (through the Einstein equation) and
implicit (through the duality) descriptions
will help illustrating the generality of this problem and toward 
finding the explicit solutions.

\section{The $h$-problem}

In DBI inflation, the potential is steep so the inflaton travels near
the speed-limit provided by the warped space. For (\ref{hAdS}), the
leading behavior of the inflaton is
\bea
r \approx \pm \frac{R^2}{t} ~, ~~~~ |t| > H^{-1} ~.
\label{AdSSL}
\eea
In order to have DBI inflation, we need $\Delta t > H^{-1}$, so
the warp factor $h$ has to be smaller than $h_0 =HR$. 
This statement is model-independent once (\ref{hAdS}) is given.
To have
$N_e$ e-folds of inflation, for the IR DBI model \cite{Chen:2004gc}, 
the warp factor has
to be as small as $HR/N_e$ where $H$ is a constant; for the UV DBI
model \cite{Silverstein:2003hf}, 
as small as $HR/p$ where $p \gg 1$ is a constant and $H$ slowly drops
by a factor of $e^{N_e/p}$ during the inflation. 

Let us look at the Einstein equation for the warp factor. 
In order to have a stabilized warped compactification, we
need the source for the warping as well as some bulk fields that
stabilize the extra dimensions, although their roles can be mixed. The
situation for a full string-theoretic construction can be quite
complicated \cite{Giddings:2001yu}. 
To illustrate our main point, we first use a simpler toy model,
i.e.~the Randall-Sundrum (RS) model \cite{Randall:1999ee} 
with the Goldberger-Wise (GW) stabilization
\cite{Goldberger:1999uk,DeWolfe:1999cp}.
We shall turn to the type IIB string theory shortly. 
The IR and UV branes separated in the fifth dimension together with 
the bulk potential 
$V(\Phi)$ provide the source for the
warped space; while the GW scalar field $\Phi(r)$ 
in the bulk stabilizes the size of the extra dimension. The Einstein
equation for the warp factor $h(r)$ in (\ref{5dmetric}) is
\bea
h'^2 - H^2 h^{-2} = 
\frac{1}{24 M_5^3} \left[ \half h^2 \Phi'^2 - V(\Phi) \right] ~,
\label{heom}
\eea
where the Hubble parameter $H$ is approximately a constant (which is
true at least for a few efolds in a sustained period of inflation), 
$M_5$ is
the 5d Planck mass and the prime $'$ denotes the derivative with
respect to $r$.

We first consider the case where the bulk field contribution on the
right hand side of (\ref{heom}) to $h(r)$ is
negligible and the bulk potential is a negative constant,
$V=-24M_5^3/R^2$. 
It is not difficult to see that the term $H^2h^{-2}$ becomes
increasingly important towards the IR side, 
and the naive solution
(\ref{hAdS}) is deformed near $h\approx HR$.
The solution becomes
\bea
h = \left( \frac{r^2}{R^2} - H^2 R^2 \right)^{1/2} ~.
\label{hdeform}
\eea
This solution has been worked out before \cite{Kaloper:1999sm,DeWolfe:1999cp,Kim:1999ja}, 
and the form we present here is
related to them by simple change of variables.
The singularity at the horizon $r=HR^2$ is a coordinate singularity.

Interestingly, the solution
(\ref{hdeform}) is related to the static AdS$_5$,
\bea
ds^2 = \frac{\tilde r^2}{R^2} (-d\tilde t^2 + d {\bf x}^2) + 
\frac{R^2}{\tilde r^2} d\tilde r^2 ~,
\eea
by a coordinate transformation
\cite{DeWolfe:1999cp,Buchel:2002wf},
\bea
\tilde t &=& -\frac{H^{-1} e^{-Ht}}{\sqrt{1-\frac{H^2R^4}{r^2}}} ~, \\
\tilde r &=& R^2 H  e^{Ht} \sqrt{\frac{r^2}{H^2R^4}-1} ~.
\eea
This observation makes it straightforward to generalize the metric
(\ref{hdeform}) to the case of the type IIB string theory, by adding a
trivial angular part $R^2 d\Omega_5^2$ and a correspondingly
transformed RR 4-form \cite{Buchel:2002wf},
\bea
{\cal C}_{\it 4} = \left( \int \frac{4}{R} h^3 dr \right) 
d{\rm Vol}_{{\rm dS}_4} ~.
\label{RRdeform}
\eea
At this point, any other contributions from the bulk are ignored as
before.

In this case, instead of working with the naive form of the metric
(\ref{hAdS}), the relevant region for the DBI inflation is given by
(\ref{hdeform}). The speed-limit still approaches zero near the new
horizon $r=HR^2$, but it has a
very different behavior. For example, under
a generic potential $V=V_0- \half
\beta H^2 r^2$ ($\beta \sim 1$), the inflaton behaves as
\bea
\frac{r}{HR^2} -1  = \frac{2}{e^{-2Ht}-1} 
- \frac{200}{\beta^2} ~e^{6Ht} + \cdots~, ~~~~~ t<-H^{-1} ~,
\label{InflatonBeh}
\eea
where the first term gives the speed of light,
and the second term is determined by the
RR-potential (\ref{RRdeform}) and $V$.\footnote
{
To derive this solution from the equation of motion
\bea
\frac{d}{dt} \left( \frac{h^2 \dot r}{\sqrt{h^4-\dot r}} \right)
+3H \frac{h^2 \dot r} {\sqrt{h^4-\dot r}}
+ \frac{\partial_r(h^2) (2h^4-\dot r^2)}{\sqrt{h^4-\dot r}}
-\partial_r C_4 +\partial_r V =0~,
\label{eom_r}
\eea
one can expand $r(t)$ around the speed of light,
$r=HR^2+2HR^2/(e^{-2Ht}-1) + a HR^2 e^{b H t} + \cdots$, $t<-1/H$, 
in the region
$\frac{r}{HR^2} -1 \ll 1$ where the speed limit becomes important and where
$C_4 \to \frac{16\sqrt{2}}{5} H^4R^4(\frac{r}{HR^2}-1)^{5/2}$. One can
see that the 2nd, 3rd and 5th terms in Eq.~(\ref{eom_r}) are important
and the coefficients $a$ and $b$ in the above ansatz can be
determined. One can also see that the exponential dependence on $t$ in
the expansion is due to the fact that the speed of light exponentially
approaches the horizon, and this is quite generic against the shape of the
potential. For example adding a linear potential
makes $\partial_r V$ a
different constant near $r\to HR^2$, and this will only change the
parameter $\beta$ in (\ref{InflatonBeh}).
}
Notice that the 
inflaton now approaches the
horizon exponentially fast with $|t|$, in contrast to the inverse
relation in Eq.~(\ref{AdSSL}). This leads to the rapid exponential
growth of the Lorentz factor
\bea
\gamma \approx \frac{\beta}{20} e^{2 N_e} ~,
\label{gammadeform}
\eea
where $N_e \approx H |t|$ is the number of e-folds to the end of
inflation. This is in contrast to the much milder growth behavior 
(such as
linear \cite{Chen:2004gc}) in the AdS geometry (\ref{hAdS}).
Now the main problem is the rapid growth of the probe
backreaction. In terms of brane inflation in warped compactification,
to support an ultra-relativistic probe brane with Lorentz factor
$\gamma$, the warped throat needs to have a charge $\gg
\gamma$ \cite{Silverstein:2003hf,Chen:2004hua}. 
To have 60 e-folds, the growth in (\ref{gammadeform}) is
undesirably large.\footnote{Because the magnitude of the
  non-Gaussianity is given by the estimator $f_{NL} \sim \gamma^2$,
  Eq.~(\ref{gammadeform}) would seem to
  certainly violate the observational bound. However this
  conclusion may be too naive since the stringy effects are rapidly
  becoming important due to the warping and the relativistic effect,
  so that the field-theoretic
  results for the primordial fluctuations may not apply in the relevant
  window \cite{Chen:2005ad,Bean:2007eh}. Here we stick to the
  model consistency conditions before any comparison with
  observational data is made.}

Since the second term on the left hand side of (\ref{heom}) is always
present, we
regard the above problem to be a generic feature. As we will discuss
in the following sections, in order to have successful DBI inflation, 
other contributions
such as the non-renormalizable operators or
bulk fields have to be introduced to cancel this term.

The presence of the GW bulk field can also modify the term $h$, 
and therefore gives a different warping behavior from
(\ref{hAdS}) even in the absence of the inflationary background.
At least for the toy model, this does not improve the
situation for a more general class of
warping. Consider a different warp factor
\bea
h=\left( \frac{r}{R} \right)^\alpha ~.
\eea
Setting $H=0$ one can obtain what is needed for the bulk and boundary
potentials for the $\Phi$ field, by solving (\ref{heom}) and another
equation of motion for $\Phi$. We ignore such details here. The point
here is that once such a warping is obtained, adding back the term with
$H$, the deformation happens for $h'^2 < H^2 h^{-2}$, i.e.
\bea
h < \left( \frac{HR}{\alpha} \right)^{\frac{\alpha}{2\alpha-1}} ~.
\label{hcond1}
\eea
The speed-limit of the inflaton is
\bea
r = \left[ \pm \frac{R^{2\alpha}}{(2\alpha-1)t}
  \right]^{\frac{1}{2\alpha-1}} ~,
\eea
so in order to have DBI inflation,
\bea
h < \left( \frac{HR}{2\alpha-1} \right)^{\frac{\alpha}{2\alpha-1}} ~.
\label{hcond2}
\eea
Note that the different geometry generically changes the critical
value $h_0$. However the position of the backreaction deformation is
also changed. To have
(\ref{hcond1}) much smaller than (\ref{hcond2}), we need $\alpha$ to be
very close to $1/2$. But in such a case the speed of light approaches
the horizon near exponentially with $t$, 
and this will make the inflaton Lorenz factor
grow rapidly (close to exponentially) and lead to a very large probe
backreaction as we considered previously.

\section{Relation to the $\eta$-problem}

In the spirit of Maldacena's conjecture \cite{Maldacena:1997re}, 
in the type IIB string theory, 
a warped space with AdS$_5$ $\times$
$S^5$ attached
to a stabilized six dimensional bulk is dual to a strongly coupled
${\cal N}= 4$ SYM field theory coupled to gravity. Aspects of this
duality are discussed in Ref.~\cite{Gubser:1999vj,ArkaniHamed:2000ds}. 
For our interest, we impose an inflationary background on both sides
of the duality and study the properties of fields in this background.
We study a dilaton field
$\phi$ in the deformed AdS space in the higher dimensional
string theory and then inspect what it means for the dual lower
dimensional field theory.

The equation of motion for the s-wave dilaton,
\bea
\frac{1}{\sqrt{-G}} \partial_M ( \sqrt{-G} G^{MN} \partial_N \phi)=0
~,
\eea
becomes
\bea
\frac{\partial^2 \phi}{\partial t^2} + 3H \frac{\partial \phi}{\partial t}
- \frac{\nabla^2}{a^2} \phi - \frac{1}{h} \frac{\partial}{\partial r}
\left( h^5 \frac{\partial}{\partial r} \phi \right) =0 ~.
\eea
Separating the variables,
\bea
\phi(t,r) = \varphi(r) g(\bx,t) ~,
\eea
we get two equations,
\bea
\frac{1}{h} \frac{d}{dr} 
\left( h^5 \frac{d}{dr} \varphi \right)
+ m^2 \varphi &=& 0 ~, 
\label{r_comp} \\
\ddot g + 3H \dot g - \frac{\nabla^2}{a^2} g + m^2 g &=& 0 ~,
\label{t_comp}
\eea
where $m^2$ is the eigenvalue of the first differential
equation.
Redefining 
\bea
y^2\equiv \frac{r^2}{H^2 R^4} -1 ~,
\eea
Eq.~(\ref{r_comp}) has only one dimensionless parameter $m^2/H^2$,
\bea
y^2(1+y^2) \frac{d^2\varphi}{dy^2} + y(5y^2 + 4) \frac{d\varphi}{dy} +
\frac{m^2}{H^2} \varphi =0 ~.
\label{varphieom}
\eea
Starting with the normalizable solution $\varphi \sim 1/y^4$ at $y\gg
1$,
approaching $y \ll 1$, $\varphi$ generally behaves as
\bea
\varphi = c_+ y^{n_+} + c_- y^{n_-} ~.
\label{varphisolution}
\eea
The two linearly independent
solutions are given by 
\bea
n_{\pm} = -\frac{3}{2} \pm
\sqrt{\frac{9}{4}-\frac{m^2}{H^2}} ~,
\eea
where $n_{\pm}$ can be complex ($\varphi$ is real).
First we can see that $m^2/H^2 \le 0$ is unacceptable. Multiplying
Eq.~(\ref{varphieom}) by $y^2 \varphi/\sqrt{1+y^2}$ and integrating
over $y$, we get
\bea
\left[ y^4 \sqrt{1+y^2} \varphi \frac{d\varphi}{dy} \right]_{y_m}^\infty
+ \int_{y_m}^\infty dy 
\left[ \frac{m^2}{H^2} \frac{y^2}{\sqrt{1+y^2}} \varphi^2
- y^4 \sqrt{1+y^2} \left( \frac{d\varphi}{dy} \right)^2 \right] =0 ~,
\label{varphicond}
\eea
where we denote the lower limit as $y_m$.
For the normalizable and regular solution,
the first term above vanishes and this equation cannot be satisfied if
$m^2\le 0$.

If we trust (\ref{hdeform}) all the way to $r=HR^2$, both solutions
are irregular at the horizon.\footnote{
In this case, even if we only look for normalizable solutions, we
still cannot find the eigenvalue due to the singular behavior of the
solutions at the horizon. 
To be normalizable near $y=0$, the metric (\ref{hdeform}) 
requires $\varphi\sim y^n$ with $n>-3/2$. This leaves a possible
region $0<m^2/H^2<9/4$ for the branch $\varphi \sim y^{n_+}$. However
numerical calculation suggests no eigenvalue due to the rapid growth
(as $y \to 0$) of the other non-normalizable branch.
}
In fact there should be a
smoothing cutoff at a small $y_m$, because the local blue-shifted
Hubble parameter
is infinite at $y=0$ and particle creation will become important.
The regular solution is possible with this cutoff by requiring
$d\varphi/dy|_{y_m}=0$. This can be satisfied for $m^2/H^2>9/4$ where
the solution starts to oscillate. The periodicity of the
trigonometric function determines the spacing between the discrete
eigenvalues to be roughly 
$\Delta \sqrt{m^2/H^2-9/4} \sim \pi/|\ln y_m|$.
Indeed, numerical calculation finds, for example for $y_m=0.01$, 
that $m^2/H^2 = 2.683, 3.896, 5.825, \cdots$.
So we have shown that the eigenvalue $m^2$
has a positive gap of order $H^2$.

Let us look at the second equation (\ref{t_comp}). This is the familiar
equation of motion for a scalar field 
with mass $m$ in a dS$_4$ background. 
Now we can come back to the duality conjecture
mentioned at the beginning of this section.
The dual 4d particle state can be thought of as having a normalizable 
wave-function
in the direction of the AdS space from the dual higher dimensional point
of view. The specific deformation of this wave-function at the IR end
of the AdS space is translated into
an effective mass gap of order $H$ from the 4d field
theory point of view.
Qualitatively the appearance of this mass gap is the origin of the
field theoretic $\eta$-problem in the slow-roll inflation. 
For any candidate scalar
inflaton, such a mass gives contribution of order unity to
the slow-roll parameter $\eta=\mpl^2 V''/V$.

The development of the mass gap is due to the IR cutoff near the
horizon $r=HR^2$. Otherwise, the IR direction of the AdS space is
effectively non-compact for the wave-function, and the 4d field theory
is scale-invariant.
This is qualitatively similar to what happens
in the confinement and finite temperature theories
\cite{Witten:1998zw}.
This is also qualitatively similar to the general argument of the mass
spectrum of the KK particles in such a warped space with IR cutoff at
$h_0$,
i.e.~the spectrum is discretized in the unit $h_0/R$ where $h_0$ is now
$HR$.

For a comparison,
Ref.~\cite{McAllister:2007bg} (P26) discussed a fine-tuning issue
that has some interesting relation to the arguments in this section,
but with a different physical origin. There one
starts by assuming a mass gap
$m\sim F/\mpl$ through gravity mediation 
($F$ is the supersymmetry breaking F-term) on
the field theory side, which in principle does not have to be the same
as our $H$ but may be regarded analogously as our $\eta$-problem. Then
this mass gap should be dual to a
cutoff in the AdS geometry by running our argument reversely (although
without an explicit metric).
The starting mass gap 
is in the strongly coupled field theory side and is
not explicit so far. In addition, Ref.~\cite{McAllister:2007bg} uses
a chaotic potential
with the same mass $m\sim F/\mpl$ for the probe brane through gravity
mediation. This is
a special case, since for the chaotic potential $m^2 \phi^2$ the
DBI inflation requires $m \gg
\mpl/\sqrt{N}$ (where $N$ is the effective throat
charge) \cite{Silverstein:2003hf}, hence no specific
relation to $m$ defined above.
The consistency criteria used to see the problem is that the potential
energy dominates
over the kinetic energy. Although for the UV model, this leads to a
tuning problem, it will be a very weak constraint for the IR
DBI model (since the potential there is dominated by a constant) and
therefore will not lead to a fine-tuning problem.
In this paper, we work on the string theory side first
by identifying the source term of the backreaction in the Einstein
equation and examining the explicit form of the deformed metric. 
The criteria used is the universal speed-limit (\ref{AdSSL}).
We saw
that the source of the $h$-problem is solely due to the inflationary
background with the Hubble parameter $H$,
and is independent of specific models and 
of whether the probe brane is coupled, or the mass gap is generated,
through the gravity or gauge mediation.
(In our point of view, 
such model dependent aspects in Ref.~\cite{McAllister:2007bg} may
actually become important as
the counter-sources used to tune away the $h$-problem discussed here.)
So the fine-tuning issue in the $h$-problem
is more explicit and model-independent. Here the corresponding mass gap in
the field theory side is only implied via the duality.

There is also another aspect in which the $h$ and $\eta$-problems
are related. We introduce a probe brane, namely a candidate inflaton, in
this geometry. We have seen in the previous section
that the DBI inflation is spoiled. We now consider the probe brane moving
much slower than the speed-limit and see if it can satisfy the slow-roll
condition. In this case, one can perturbatively expand the DBI action
(\ref{DBIkinetic}). The interesting region is now $r\gg HR^2$. 
Although in
the leading order the deformed gravity field (\ref{hdeform}) and
the deformed RR field (\ref{RRdeform}) cancel, at the subleading order
they do not. The net
contribution is a mass term for the probe brane, $m^2=H^2$
\cite{Buchel:2003qj}. This should be viewed as 
the dual string theory description
\cite{Buchel:2003qj} of the
conformal coupling in the field theory description of the KKLMMT
slow-roll inflation
\cite{Kachru:2003sx}. So the same deformed geometry causes both the
$h$-problem and $\eta$-problem, this time both in the string theory
side.

\section{Tuning}

The full deformation of the metric
due to the inflation 
should be more subtle than that due to the AdS blackhole.
In light of
the AdS/CFT duality relation discussed above, we can think of this
problem in the 4d field theory side.
In the AdS blackhole case \cite{Witten:1998zw}, 
the dual field theory has a finite temperature $T$. One
can fix $T$ while sending the cutoff scale to
infinity. In other words, the correction to the mass gap from the
non-renormalizable operators, i.e.~those irrelevant operators suppressed
by the UV cutoff 
scale such as $\mpl$, will be suppressed by $\mpl$. 
In the inflationary case we cannot do that since $H$ itself is
suppressed by the Planck mass $\mpl$ (typically with the inflationary
energy scale or supersymmetry breaking scale holding fixed). Therefore on
the field theory side
various non-renormalizable operators may provide an effective
mass-squared of the same
order of magnitude $H^2$, but with an opposite sign. 
From the discussion below Eq.~(\ref{AdSSL}),
a reduction of the mass gap from the order $H$ to $0.01H$ on the field
theory side will open up a portion of warped 
space on the string theory side for $100$ e-folds of DBI inflation.
This is interestingly similar amount of tuning as for the slow-roll
potential.
In terms of the warped compactification, 
this tuning depends on
the details of the UV completion, namely
the way in which the warped space is attached to the bulk and the
physics in the bulk such as the supersymmetry breaking. For example,
such non-renormalizable
operators are also expected to be the generic
tuning source for the slow-roll inflationary potential.
Constructing explicit examples will be a very interesting open
question.

Another related way to provide the
tuning is to arrange the bulk field contribution
on the right hand side of the equation of motion (\ref{heom}) 
to cancel the second term with $H$
on the left hand side. 
This amounts to for example setting up the potentials for the bulk
field. This approach is more feasible
in more complicated examples instead of this toy model.

To conclude, we have shown that, in the DBI inflationary mechanism, 
there is a
fine-tuning $h$-problem concerning the explicit construction of the
warped
space taking into account of the backreaction from the inflationary
background. The non-renormalizable
operators in the dual field theory and the bulk fields should play
important roles in such fine-tuning. Without
tuning, the part of the naive warped space required for the DBI
inflation will be deformed by the inflationary background through
Einstein's equation in the extra dimensions. In the
deformed geometry, the probe backreaction of the inflaton
will spoil the DBI inflation
unless the background throat charge is undesirably large. This
resembles the familiar fine-tuning $\eta$-problem in the slow-roll
inflationary mechanism concerning the explicit construction of the
flat potential taking
into account of the backreaction from the inflationary background. The
origins and tunings of these two problems are interestingly 
related through the
AdS/CFT relation. Nonetheless, these two inflationary mechanisms have
distinctive observational predictions, and should reveal different
aspects of the underlying fundamental theory \cite{Chen:2005fe,Shiu:2006kj,Bean:2007hc,Bean:2007eh,Thomas:2007sj,Alabidi:2008ej,Lorenz:2008et}.

\medskip
\section*{Acknowledgments}

I would like to thank Allan Adams, Andrew Frey, Shamit Kachru, Justin
Khoury, Louis Leblond, John McGreevy, Wati
Taylor, and especially Hong Liu, Juan Maldacena, Fernando Quevedo
for helpful discussions, and Eva Silverstein for explaining the discussions
in Ref.~\cite{McAllister:2007bg} related to our Sec.3.
This work is supported by the US Department of Energy under
cooperative research agreement DEFG02-05ER41360.


\end{document}